\documentclass{PoS}

\newcommand{\ie}{{\it i.e.}}

\newcommand{\eg}{{\it e.g.}}

\newcommand{\eq}{Eq.}

\newcommand{\fig}{Fig.}

\newcommand{\Sec}{Section}

\newcommand{\mceq}{{\sc MCEq}}
\newcommand{\rmd}{{\mathrm d}}
\newcommand{\xl}{x_{\rm Lab}}

\newcommand{\equ}[1]{\eq~(\ref{eq:#1})}
\newcommand{\figu}[1]{\fig~\ref{fig:#1}}

\newcommand{\bi}{\begin{itemize}}
\newcommand{\ei}{\end{itemize}}

\usepackage{amsmath}
\usepackage{sidecap}
\usepackage{multirow}

\title{Constraints on light meson production in air-showers with atmospheric neutrinos below 1~TeV interacting in IceCube's DeepCore}

\ShortTitle{Kaon to pion ratio measurement}

\author{
The IceCube Collaboration\footnote{For collaboration list, see PoS(ICRC2019) 1177.}\\
{\itshape \href{http://icecube.wisc.edu/collaboration/authors/icrc19_icecube}{http://icecube.wisc.edu/collaboration/authors/icrc19\_icecube}}\\
E-mail: \email{anatoli.fedynitch@ualberta.ca}
}

\abstract{

The IceCube Neutrino Detector observes tens of thousands of atmospheric neutrinos every year with its low energy extension DeepCore, covering the neutrino energy range from GeV to TeV. These neutrinos can be used to study particle interactions in conditions that are inaccessible to current accelerator-based experiments. Here,  we present constraints concerning the production of light and strange mesons in atmospheric particle cascades using a dedicated analysis based on the shape of the neutrino spectrum in zenith and energy. We report the first measurement of the atmospheric K/Pi-ratio in a constrained energy range exclusively from neutrino observations.

\vspace{4mm}
{\bfseries Corresponding authors:}
\speaker{Anatoli Fedynitch}$^{1}$, Juan-Pablo Y\'{a}\~{n}ez$^{1}$\\
{$^{1}$ \itshape University of Alberta}\\

}

\FullConference{36th International Cosmic Ray Conference -ICRC2019-\\
		July 24th - August 1st, 2019\\
		Madison, WI, U.S.A.}

\begin{document}

\sidecaptionvpos{figure}{c}
\section{Introduction}
The interactions of cosmic ray nuclei with the atmosphere evolve into extensive particle cascades, in which the decays of unstable particles produce muons and neutrinos. These so called atmospheric leptons remain as the dominant particle species at ground and can be detected by surface and underground detectors. IceCube is a cubic-kilometer neutrino detector installed in the ice at the geographic South Pole \cite{Aartsen:2016nxy} between depths of 1450 m and 2450 m, completed in 2010. It detects thousands of muon events each second and tens of neutrinos per hour, roughly corresponding to one neutrino event per hour at analysis level, and hence it is a unique laboratory to study atmospheric leptons.

While the main goal of IceCube is to study astrophysical neutrino sources, the sheer number of atmospheric events constitutes a natural beam of neutrinos that makes it possible to pursue a particle physics program including measurements of standard and non-standard neutrino oscillations \cite{Aartsen:2019tjl,Aartsen:2017bap,Aartsen:2017xtt}. These measurements rely on a theoretical characterization of the energy and angular dependence of the atmospheric neutrino flux \cite{Gaisser:2002jj} in the absence of oscillations and matter effects. Theoretical uncertainties in flux calculations can impact or bias the determination of fundamental neutrino properties, constituting a significant systematic uncertainty.

In this paper, we use a sample of atmospheric neutrinos collected with the densely instrumented DeepCore subarray that is embedded in the part of IceCube with the best ice properties and instrumented with high-quantum efficiency photomultiplier tubes. The surrounding standard IceCube strings are used as a veto for down-going atmospheric muons, resulting in a full sky coverage in an energy range from a few GeV up to 180 GeV. We exploit the impact of the hadronic particle production on the shape (in the energy-zenith plane) of the observed atmospheric neutrino flux to infer one of the leading uncertainties for inclusive lepton flux (integrated over cosmic ray energy) calculations - the ratio of kaon to pion multiplicities in the fragmentation region \cite{Barr:2004br}, \textit{i.e.} at large values of Feynman $x_\text{F}$, at equivalent projectile energies above that of modern fixed target facilities.

\section{Spectrum-weighted moments}
\label{sec:zfacs}

The modeling of hadronic interactions induced by primary cosmic rays and their secondary particles is the largest source of uncertainty in atmospheric neutrino computations \cite{Barr:2004br}. In longitudinal (1D) cascade calculations, particle production is described in terms of the inclusive multiplicity spectra
\begin{equation}
\label{eq:diff_cs}
    \frac{\text{d}N_{A \to h}}{\text{d}E_h}(E_\text{CR},A_\text{CR}, E_h) = \frac{1}{\sigma_\text{prod,A-air}(E_\text{CR})} \frac{\text{d}\sigma_{A \to h}}{\text{d}E_h}(E_\text{CR}, A_\text{CR}, E_h),
\end{equation}
where the index $h$ denotes the hadron type, \ie{} pion, kaon, proton, etc., $A_\text{CR}$ and $E_\text{CR}$ are the mass and the energy of the cosmic ray nucleus, and $E_h$ is the energy of the hadron, $\sigma_\text{prod}$ the interaction cross section and $\frac{\text{d}\sigma_{A \to h}}{\text{d}E_h}(E_\text{CR}, A_\text{CR}, E_h)$ the differential particle production cross section. Semi-analytical solutions for the inclusive flux $\Phi$ of leptons $\ell$ in the atmosphere are explained in detail in \eg{} \cite{Gaisser:2016uoy} and also in a different contribution at this conference \cite{Gaisser:2019icrc}. The solutions are of the form  
\begin{equation}
\label{eq:semi-anal}
\Phi_\ell(E) = \frac{\phi_\text{N}(E)}{1 - Z_{\rm NN}}\sum_{\substack{h = \pi,\\ \text{K}, \text{K}^0_\text{L},\dots}} \frac{Z_{\text{N}h,\gamma} Z_{h \to \ell,\gamma}}{1 + B_{h\ell} E \cos\theta/\varepsilon_h},
\end{equation}
with a quantity $B_{h\ell}$, that depends on the spectral index of cosmic rays and the decay kinematics of the meson $h$. The so called energy-independent spectrum weighted moments (Z-factors)
\begin{equation}
\label{eq:z-factor}
  Z_{Nh} = \int_0^1 \rmd{} \xl ~ \xl^{\gamma - 1} \frac{\rmd N_{\text{N} \to h}}{\rmd \xl},
\end{equation}
take into account the multiplicity and the spectrum of the secondary meson $h$ into account. Similarly, the decay Z-factor contains the branching ratio of $h$ into leptons of type $\ell$. In this energy-independent approach, one assumptions the scaling given in differential spectra \equ{diff_cs}, a weak energy-dependence of the interaction cross section, and a constant power-law index $\gamma$ of the primary cosmic ray flux at the top of the atmosphere. In that case it is sufficient to use the fractional energy $x_\text{Lab} = E_\text{h}/E_\text{projectile}$ in place of two-dimensional distributions in projectile and secondary energy. The critical energies $\varepsilon_h$ approximately describe the energy at which the interaction probability becomes higher than the decay probability. This leads to breaks in the spectra of secondaries, \ie{} $\phi_\ell \sim E_\ell^{\gamma}$ at $E_\ell  \ll \varepsilon_h$, and $\phi_\ell \sim E_\ell^{\gamma - 1}$ at high energies. Hence, the inclusive flux of neutrinos and muons follows the cosmic ray spectrum at low energies, and at the highest energies is a power steeper.

When relaxing the scaling assumptions, a more advanced form of the Z-factor can be written down
\begin{equation}
\label{eq:z-factor-edep}
  Z_{Nh}(E_\text{pr}) = \int_0^1 \rmd{} \xl ~ \xl^{\gamma(E_\text{pr}) - 1} \frac{\rmd N_{\text{N} \to h}}{\rmd \xl}(E_\text{pr}).
\end{equation}

Note, that this definition is different from the so-called energy-dependent Z-factor introduced in \cite{thunman_1996}.
\begin{SCfigure}[0.6]
\centering
\includegraphics[width=0.65\textwidth]{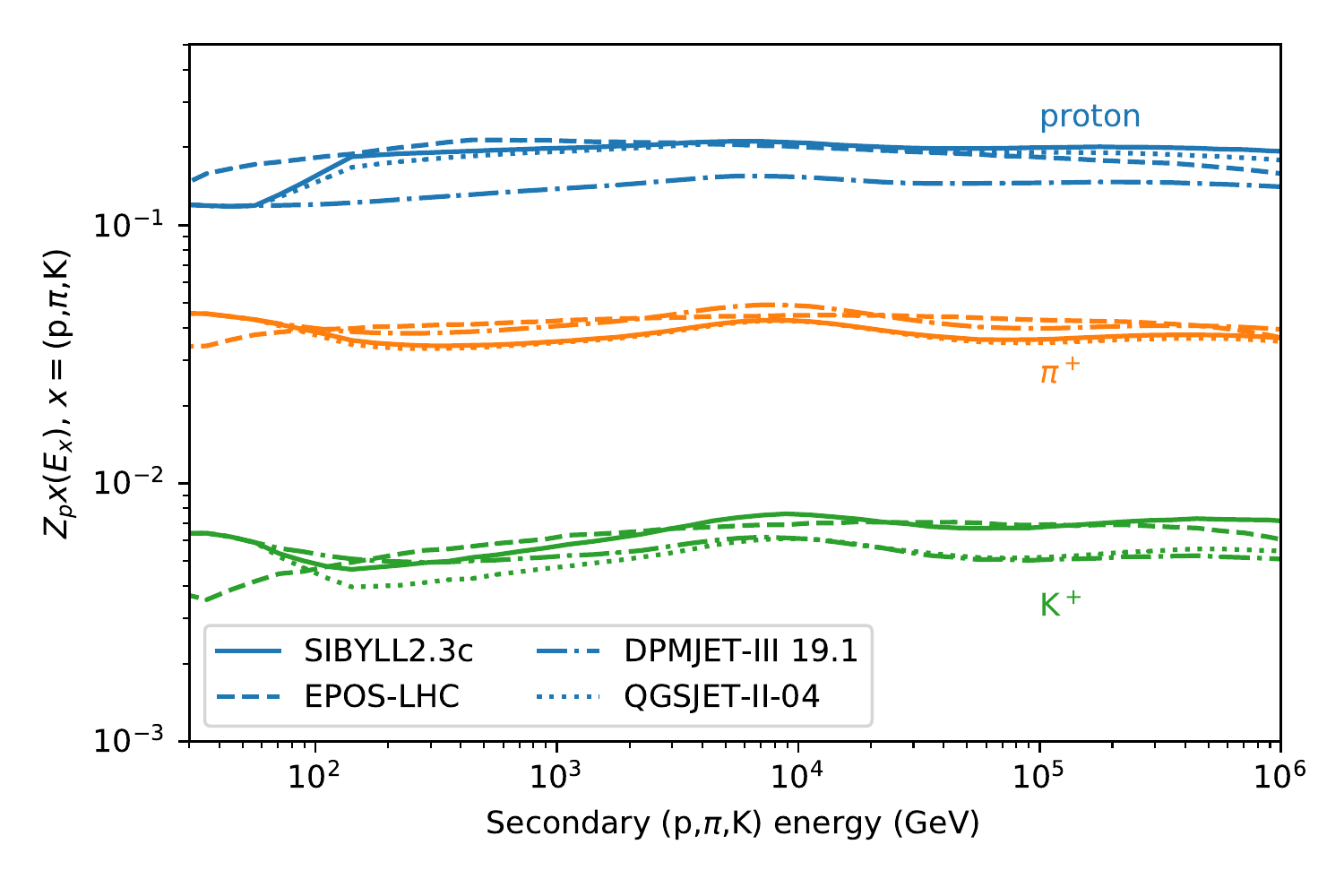} 
\caption{Energy dependence of Z-factors calculated according to \equ{z-factor-edep} for a single choice of the primary cosmic ray spectrum {\it GSF} from \cite{Dembinski:2017zsh}. The energy dependence originates from non-(Feynman-)-scaling of the interaction models, the cross section and the spectral index of the cosmic ray nucleon flux. Note Z-factors are independent of the atmospheric leptons flavor.}
\label{fig:edep-zfac}
\end{SCfigure}
Figure \ref{fig:edep-zfac} demonstrates the energy dependence of the $Z_{\text{p}h}$ for different hadronic interaction models \cite{Riehn:2017mfm,Pierog:2013ria,Ostapchenko:2010vb,dpmjetIII,Fedynitch:2015kcn}.

For the modeling of this measurement, we will use the iterative cascade equation solver \mceq{} that does not make use of the above mentioned method. However, the Z-factor framework is an essential element in the definition of the $K/\pi$ ratio.

\section{Atmospheric $K/\pi$ ratio}
\label{sec:kpi_ratio}
\begin{figure}
\centering
\includegraphics[width=0.85\textwidth]{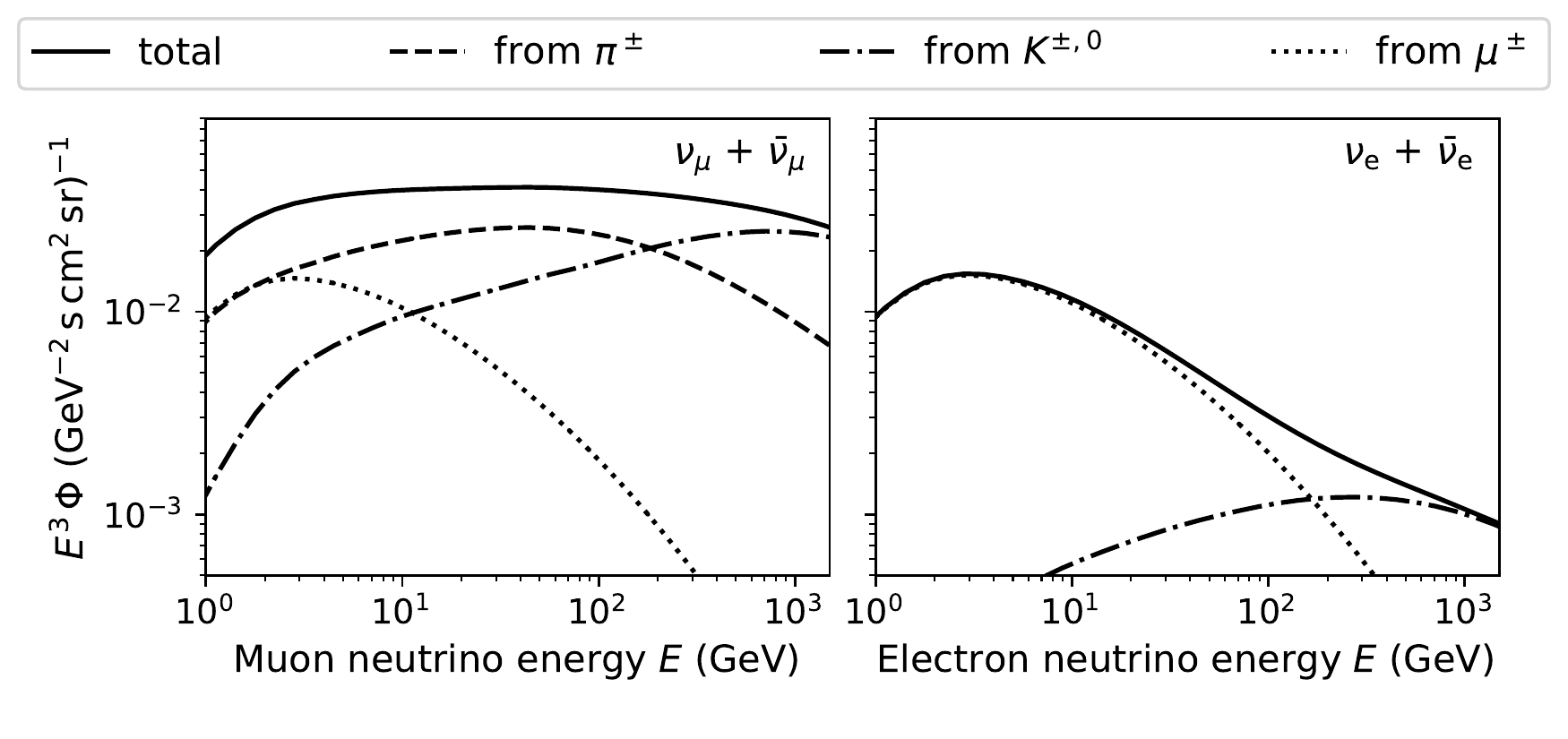} 
\caption{Composition of the atmospheric muon (left) and electron (right) neutrino flux from individual ``hadronic'' components for the zenith angle $\cos{\theta} = 0.5$. For more inclined angles the transition between pion and kaon dominated regimes in the muon neutrino flux moves to higher energies. While for electron neutrinos the contributions from direct pion decays is very small, charged pions indirectly affect the electron neutrino flux through the decay of secondary muons that predominantly come from pion decays at these energies.
\label{fig:neutrino_components}}
\end{figure}
The break-down of the atmospheric neutrino flux into different hadronic components is illustrated in \figu{neutrino_components}. Neutrinos are predominantly created in weak two-body decays of the most abundant mesons - charged pions and kaons. Despite a strongly suppressed two-body decay of pions into electron neutrinos, pions contribute to low-energy electron neutrinos via decays of muons that predominantly come from pion decays.

The shape of the inclusive neutrino flux is defined by the first few interactions at high altitudes with energies close to that of the initial cosmic ray projectiles \cite{Fedynitch:2018cbl}, in contrast to extensive air-showers where the pions observed by surface detectors originate from low-energy interactions of secondary pions. The method developed by \cite{Grashorn:2009ey} makes use of this fact and relates the seasonal variations of the muon rate observed by a deep underground detector to the quantity
\begin{equation}
\label{eq:kpi-ratio}
    r_{\text{K}/\pi} = \frac{Z_\text{NK}}{Z_{\text{N}\pi}},
\end{equation}
that is proportional to the ratio of relative abundances of pions and kaons in the first few cascade generations. This $\text{K}/\pi$ ratio refers to the secondaries of the nucleon interaction with the atmosphere, hence it is independent of the lepton flavor (muons or neutrinos) in which it is measured. The measurements based on seasonal variations exploit the differences in the decay time of the secondary mesons, \ie{} the ratio between interaction and absorption probability that varies according to air density fluctuations.

\section{Neutrinos in IceCube DeepCore}\label{sec:atmnu}

The DeepCore subarray as defined in this analysis includes 8 densely instrumented strings optimized for low energies plus 7 adjacent standard strings. All details about the event selection and the event reconstruction can be found in \cite{Aartsen:2017nmd,Aartsen:2019tjl} (referred to as Sample B). For the present analysis the sample was extended up to a reconstructed energy of 180~GeV to include more atmospheric neutrinos that originate from kaon decays.

Events are classified as cascade and track-like using the likelihood ratio of reconstructions with and without a muon present. The cascade channel contains interactions from all flavors, while the track channel is dominated by charged current $\nu_\mu$ interactions. Event counts of each channel are binned in reconstructed energy (from 5~GeV to 180~GeV) and arrival zenith angle (full sky). Systematic uncertainties are included as correction functions that modify the expectation in any given bin and are correlated across all three observables. Uncertainties that have their origins in the detector response are parameterized in a multidimensional space by means of a full re-simulation of the events after varying the detector conditions. The full list of systematic uncertainties is given in Table~\ref{table:systematics}.

\begin{table}[]
\centering
\begin{tabular}{cccc}
\hline
Parameter class                       & Description                & Function                                     & Uncertainty          \\ \hline \hline
\multirow{3}{*}{Cross section}        & Axial mass of CCRES events & From GENIE                                   &     $\sigma_\mathrm{M_A}=20\%$                 \\ \cline{2-4} 
                                      & DIS NC cross section       & Global scale                                 & $\sigma_{\sigma_\mathrm{NC}}=20\%$                 \\ \cline{2-4} 
                                      & DIS CC cross section       & $E^\gamma$                                   & $\sigma_\gamma=0.02$ \\ \hline
Background                            & Atmospheric muon rate      & Global scale                                & Ratio of two PDFs                    \\ \hline
\multirow{3}{*}{Detector performance} & DOM forward acceptance     & \multirow{3}{*}{From MC}                     &   Flat prior                    \\ \cline{2-2} \cline{4-4} 
                                      & DOM sideways acceptance    &                                              & From flasher data                     \\ \cline{2-2} \cline{4-4} 
                                      & DOM global efficiency      &                                              & $\sigma_\mathrm{eff}=10\%$                 \\ \hline
Oscillations                          & $\sin^2\theta_{23}$  and $\Delta m^2_{32}$  & $\nu$ oscillations & Flat prior \\ \hline
\end{tabular}
\caption{Sources of uncertainty considered in the analysis. The atmospheric muon rate includes additional uncorrelated shape uncertainties from two data sidebands. Details can be found in \cite{Aartsen:2017nmd}. \label{table:systematics}}
\end{table}

\section{Analysis of pion and kaon relative contributions}\label{sec:analysis}

As outlined in \Sec{\ref{sec:kpi_ratio}}, the ratio of pions to kaon abundances impacts the expected energy spectrum and angular distribution of the atmospheric neutrino spectrum. A reduction of kaons reduces the muon neutrino spectrum at high energies and vertical zenith angles. The horizontal directions remain largely unaffected. This is shown in
\begin{SCfigure}[0.6]
\centering
\includegraphics[width=0.6\textwidth]{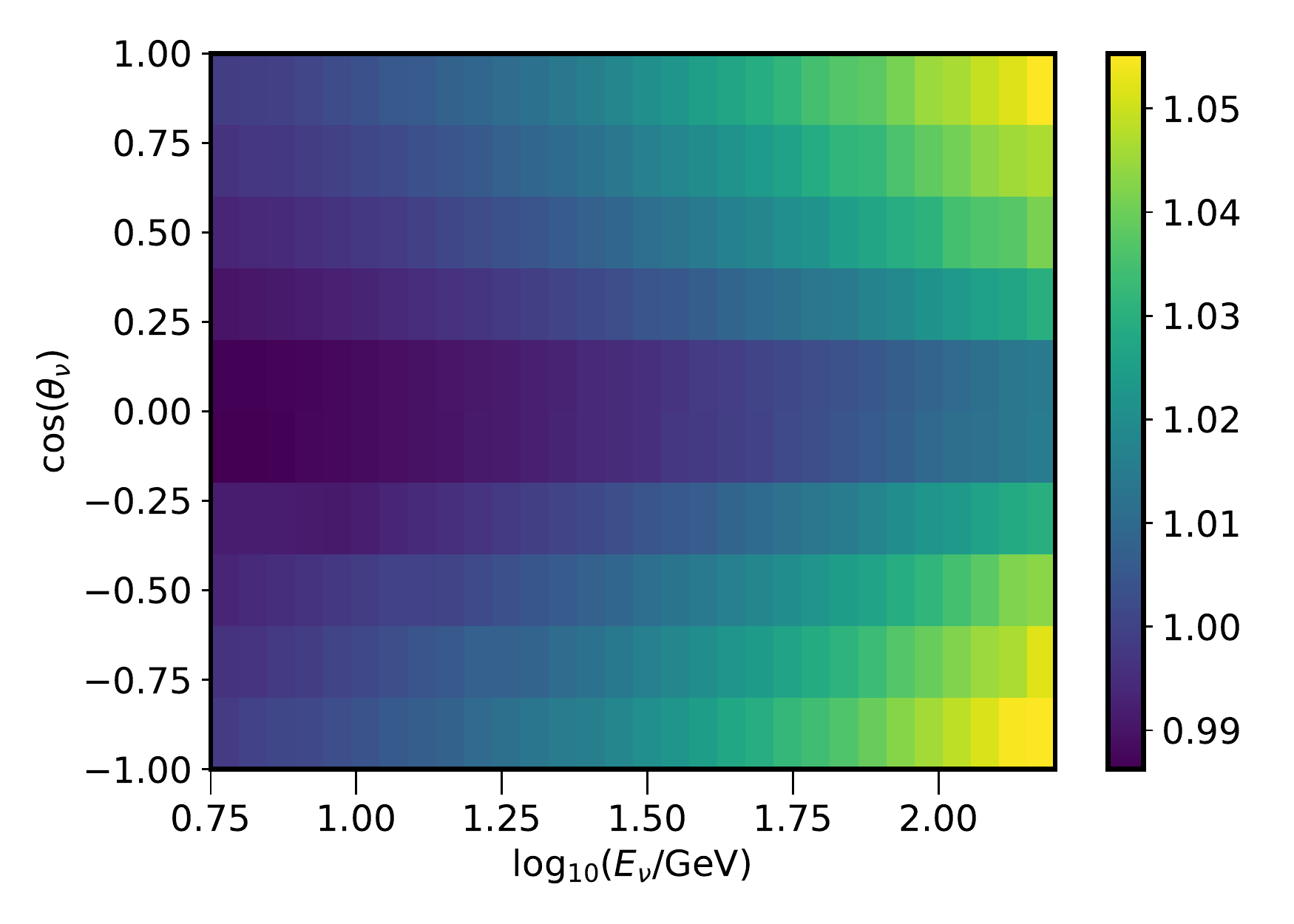} 
\caption{Relative deviation of the expected atmospheric muon neutrino flux due to an increase of pion production by 20\%. The flux in the near-vertical zenith bins increases at high energy, while the horizontal directions are unaffected.}
\label{fig:example_kaons}
\end{SCfigure}
\figu{example_kaons} that also demonstrates the quantitative impact on the effect on the  two-dimensional flux map from an increase of kaon production by 20\% at all interaction energies. For this measurement we exploit the impact on the flux shape instead of the seasonal variation method as in previous measurements \cite{Adamson:2009zf,Desiati:2011hea,Agostini:2018fnx}.

We compute the flux expectation and the correction functions (Jacobians) using the code \mceq{} \cite{Fedynitch:2015zma,Fedynitch:2018cbl,Fedynitch:2018vfe}. Using the tabulated inclusive particle spectra (\equ{diff_cs}) stored in \mceq{}, we compute the energy-dependent $Z$-factors from \equ{z-factor-edep} (see Fig.~\ref{fig:edep-zfac}) and derive a model-dependent expectation value of $r_{\text{K}/\pi}$. The two weights, $w_\pi$ and $w_k$, scale the inclusive particle production spectrum, for instance for $w_k$
\begin{equation}
\label{eq:kpi-definition}
  Z^\star_{NK} = \int_0^1 \rmd{} \xl ~ \xl^{\gamma - 1} w_k\frac{\rmd N_{\text{N} \to k}}{\rmd \xl} = w_k Z_{NK},
\end{equation}
where the impact on the spectrum and zenith distribution is handled self-consistently by \mceq{}. These parameters translate to a modified $r^\star_{K/\pi}$
\begin{equation}
    r^\star_{\text{K}/\pi} = \frac{Z^{\star}_\text{NK}}{Z^\star_{\text{N}\pi}} = \frac{w_k\,  Z_\text{NK, HI}}{w_\pi\,  Z_{\text{N}\pi,\text{HI}}} = \frac{w_k}{w_\pi}r_{\text{K}/\pi,\text{HI}},
\end{equation}
where HI denotes the specific hadronic interaction model used. Simulation studies show that the proposed re-scaling method is insufficient to fully accommodate all the differences between modern hadronic interaction models. Additional effects come from the cosmic ray model used as an initial condition for the calculation. In order to test these dependencies, we perform the measurement starting from six different neutrino flux maps, comprised of combinations of two hadronic interaction models, DPMJET III 19.1 \cite{Fedynitch:2015kcn} and SIBYLL2.3c \cite{Riehn:2017mfm}, and three different cosmic ray models; Global Spline Fit \cite{Dembinski:2017zsh}, H3a \cite{Gaisser:2012em} and GH \cite{Gaisser:2002jj}. The resulting $r^\star_{K/\pi}$ for each combination is reported together with the $\Delta \chi^2$ with respect to the global best fit. Using simulation, we verified that assuming the data to follow one model combination different from the one used in the fit, the value of $r^\star_{K/\pi}$ is retrieved within the expected errors of the measurement.

\section{Results and discussion}\label{sec:results}
\begin{figure}
\centering
\includegraphics[width=0.8\textwidth]{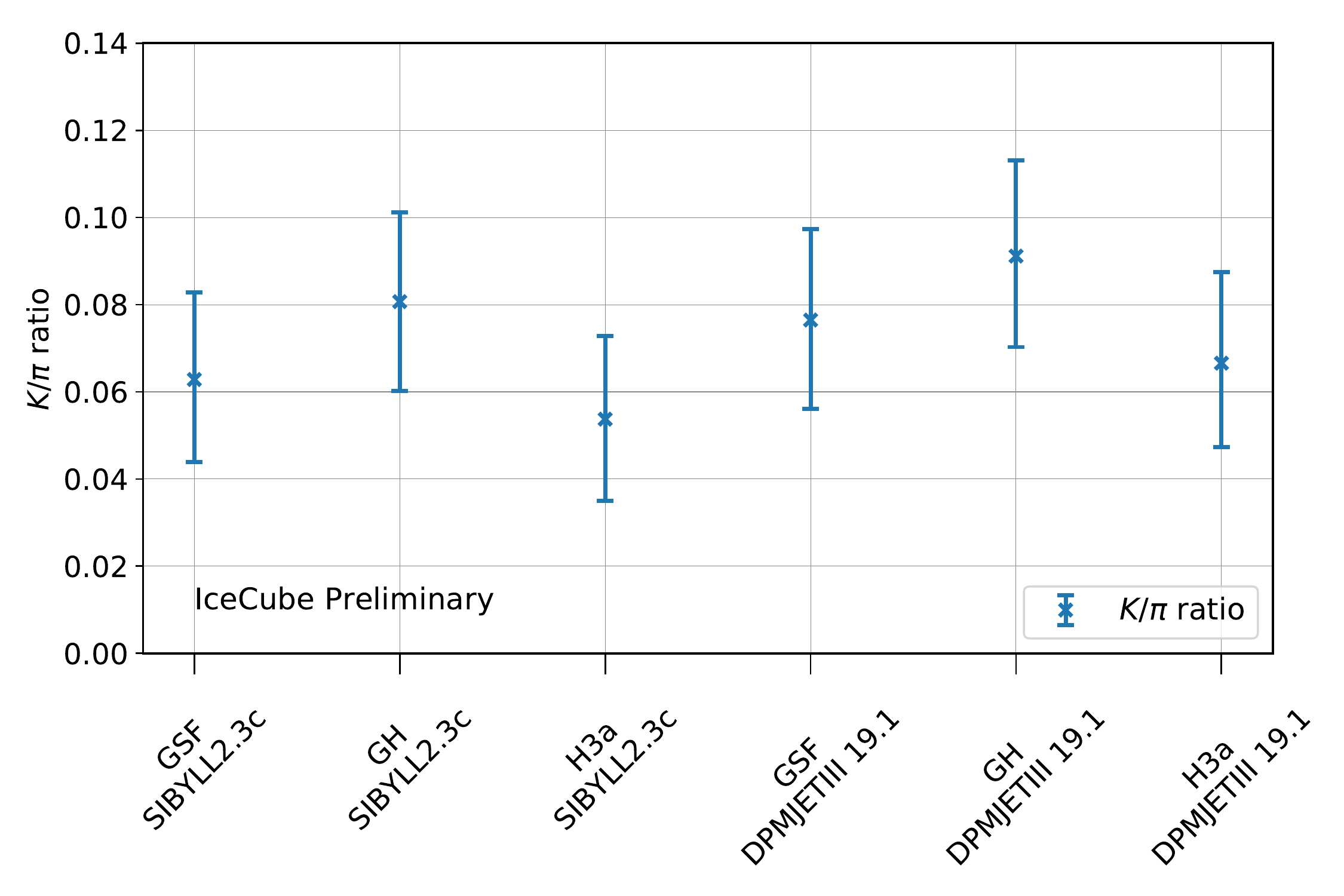} 
\caption{The best fit $r_{\text{K}/\pi}$ for each combination of cosmic ray and hadronic model. The obtained values of $r_{\text{K}/\pi}$ agree within the uncertainties of the measurement, irrespective of the model combination used as an initial assumption. The best description of the neutrino flux is achieved with the Global Spline Fit (GSF) primary flux and the {\sc DPMJET-III 19.1} interaction model.}
\label{fig:results-all}
\end{figure}
\begin{figure}
\centering
\includegraphics[width=0.8\textwidth]{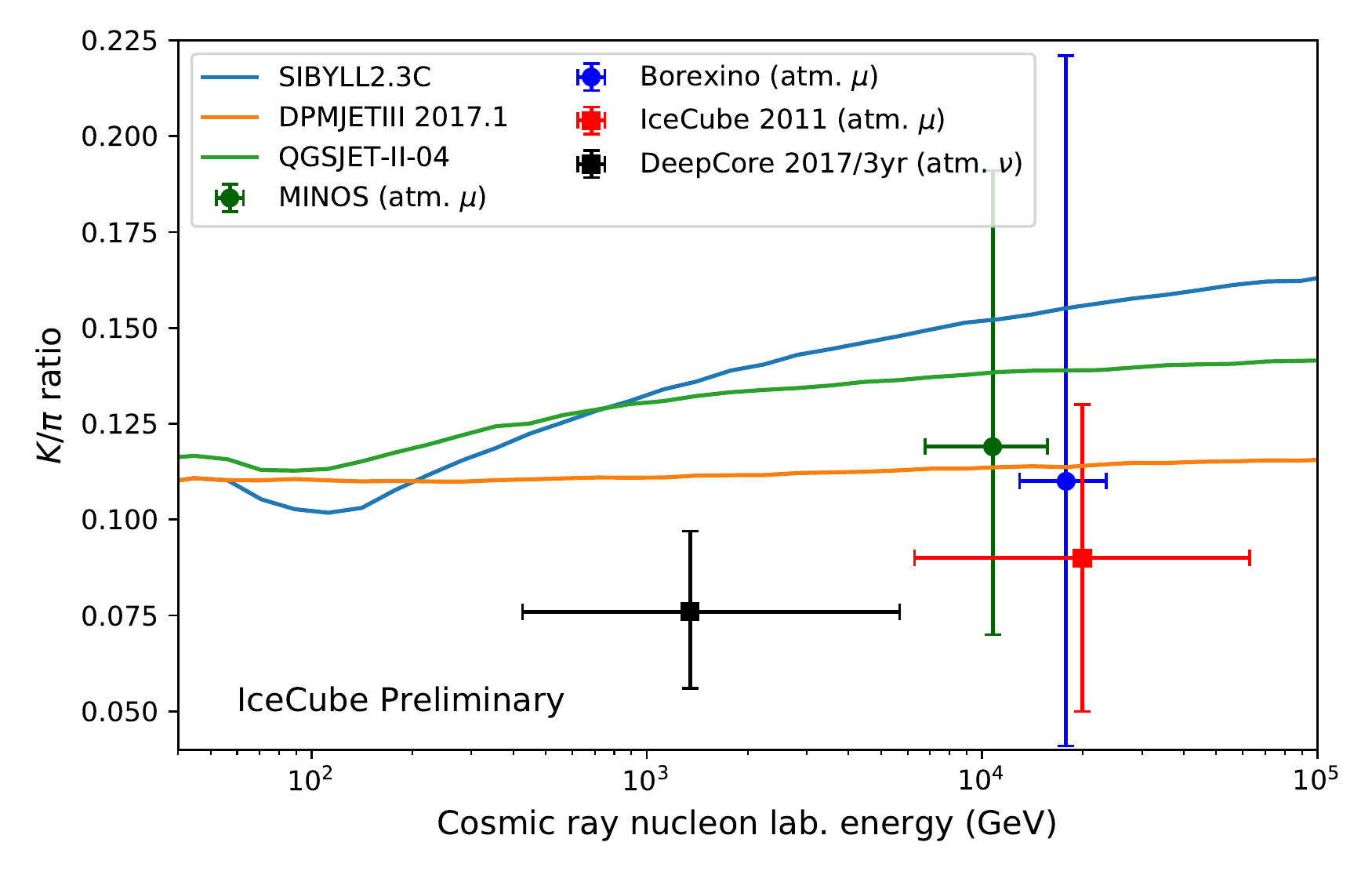} 
\caption{Measurements and theoretical expectation of the atmospheric $K/\pi$ ratio. Our result (black data point) covers a new energy range between several hundred GeV and 6~TeV and is the first result using neutrinos. The other data points from MINOS \cite{Adamson:2009zf}, IceCube \cite{Desiati:2011hea} and Borexino \cite{Agostini:2018fnx} use a method based on the seasonal variations of the atmospheric muon rate. While all measurements are compatible within their experimental errors, the tendency is for lower values compared to the theoretical expectations.}
\label{fig:results-best}
\end{figure}

The $r^\star_{\text{K}/\pi}$ results for each of the six model combinations are shown in~\figu{results-all}. The best fit to the data is obtained with $r^\star_{\text{K}/\pi} = 0.076^{+0.016}_{-0.015}$ from the combination of the Global Spline Fit and {\sc DPMJET-III 19.1}. The best fit value is consistent across all the six model combinations. 

\figu{results-best} compares our result with previous measurements that use seasonal variations of the atmospheric muon rate and the temperature correlation coefficient method \cite{Grashorn:2009ey}. The energy range covered by this measurement was determined by finding the relevant cosmic ray energies that contribute to the central 68\% of the neutrino sample. We find the relative kaon contribution to be between 1.5$\sigma$ and 2$\sigma$ lower than expected. Multiple tests were performed neglecting regions of the observable space where the simulation could be modeling the data poorly. The numerical value was found to change within the uncertainties quoted above. The study will be updated in the near future using more years of data, as well as a sample extending to higher reconstructed energy, into a region of the space where neutrinos from kaons should dominate.

\newpage
\providecommand{\href}[2]{#2}\begingroup\raggedright\endgroup

\end{document}